\documentclass[english,10pt,twocolumn,twoside]{IEEEtran}
\usepackage[T1]{fontenc}
\usepackage[latin9]{inputenc}
\usepackage{bm}
\usepackage{amsmath}
\usepackage{amsthm}
\usepackage{amssymb}
\usepackage{graphicx}

\makeatletter
\theoremstyle{plain}
\newtheorem{thm}{\protect\theoremname}
\theoremstyle{plain}
\newtheorem{prop}[thm]{\protect\propositionname}

 \usepackage{cite}

\makeatother

\usepackage{babel}
\providecommand{\propositionname}{Proposition}
\providecommand{\theoremname}{Theorem}

\begin{document}

\title{On Time-Reversal Imaging by Statistical Testing}

\author{D.~Ciuonzo,~\IEEEmembership{Senior~Member,~IEEE}\thanks{Manuscript submitted Jan. 4th, 2017; revised 28th Mar. 2017.

D. Ciuonzo is with Network Measurement and Monitoring (NM2) s.r.l.,
Naples, Italy (e-mail: domenico.ciuonzo@ieee.org).}\vspace{-0.9cm}
}
\maketitle
\begin{abstract}
This letter is focused on the design and analysis of computational
wideband time-reversal imaging algorithms, designed to be adaptive
with respect to the noise levels pertaining to the frequencies being
employed for scene probing. These algorithms are based on the concept
of cell-by-cell processing and are obtained as theoretically-founded
decision statistics for testing the hypothesis of single-scatterer
presence (absence) at a specific location. These statistics are also
validated in comparison with the maximal invariant statistic for the
proposed problem.\footnote{\emph{Notation} - Lower-case (resp. Upper-case) bold letters denote
column vectors (resp. matrices), with $a_{n}$ (resp. $A_{n,m}$)
representing the $n$-th (resp. the $(n,m)$-th) element of the vector
$\bm{a}$ (resp. matrix $\bm{A}$); $\mathbb{R}^{N}$ (resp. $\mathbb{R}^{N\times M}$)
and $\mathbb{C}^{N}$ (resp. $\mathbb{C}^{N\times M}$), are the sets
of $N$-dimensional column vectors (resp. of $N\times M$ dimensional
matrices) of real and complex numbers; $\mathbb{R}^{+}$ denotes the
set of positive real-valued numbers; $\mathcal{U}(N)$ denotes the
group of $N\times N$ unitary matrices; $\mathbb{E}\{\cdot\}$, $(\cdot)^{T}$,
$(\cdot)^{\dagger}$, $(\cdot)^{*}$, $\det\left[\cdot\right]$, $\mathrm{Tr}\left[\cdot\right]$,
$\mathrm{vec}(\cdot)$, $\left\Vert \cdot\right\Vert $ (resp. $||\cdot||_{F}$),
$\Re\{\cdot\}$ and $\Im\{\cdot\}$, denote expectation, transpose,
Hermitian, conjugate, matrix determinant and trace, matrix vectorization,
Euclidean (resp. Frobenius) norm, real and imaginary part, respectively;
$\bm{0}_{N\times M}$ (resp. $\bm{I}_{N}$) denotes the $N\times M$
null (resp. identity) matrix; $\bm{0}_{N}$ (resp. $\bm{1}_{N}$)
denotes the null (resp. ones) column vector of length $N$; $\bm{A}\otimes\bm{B}$
indicates the Kronecker product between $\bm{A}$ and $\bm{B}$; $\mathrm{diag}(\bm{v})$
denotes the diagonal matrix obtained by placing $\bm{v}$ along the
main diagonal; $\bm{x}_{1:M}$ denotes the vector obtained by concatenation
as $\bm{x}_{1:M}\triangleq\begin{bmatrix}\bm{x}_{1}^{T} & \cdots & \bm{x}_{M}^{T}\end{bmatrix}^{T}$;
$\frac{\partial f(\bm{x})}{\partial\bm{x}}$ (resp. $\frac{\partial f(\bm{x})}{\partial\bm{x}^{T}}$)
denotes the gradient of scalar-valued function $f(\bm{x})$ w.r.t.
vector $\bm{x}$ arranged in a column (resp. a row) vector; the symbols
``$\sim$'' and ``$\propto$'' mean ``distributed as'' and ``proportional
to''; $\bm{x}\sim\mathcal{C}\mathcal{N}_{N}(\bm{\mu},\bm{\Sigma})$
denotes a complex (proper) Gaussian-distributed vector $\bm{x}$ with
mean $\bm{\mu}\in\mathbb{C}^{N\times1}$ and covariance $\bm{\Sigma}\in\mathbb{H}_{N}^{++}$;
$c\sim\mathcal{C}\chi_{N}^{2}$ (resp. $c\sim\mathcal{C}\chi_{N}^{2}(\delta)$)
denotes a Random Variable (RV) distributed according to a complex
central (resp. non-central) chi-square distribution with $M$ complex
degrees of freedom (resp. with $N$ complex degrees of freedom and
noncentrality parameter $\delta$), with $c\sim\mathcal{C}\chi_{N}^{2}(\delta,a)$
representing a scaled non-central version; $f\sim\mathcal{CF}_{N,M}$
(resp. $f\sim\mathcal{C}\mathcal{F}_{N,M}(\delta)$) denotes a RV
distributed according to a complex central (resp. noncentral) F-distribution
with ($N,M$) complex degrees of freedom (resp. with ($N,M$) complex
degrees of freedom and noncentrality parameter $\delta$); $\bm{P}_{\bm{a}}$
denotes the orthogonal projection of $\bm{a}$, i.e. $\bm{P}_{\bm{a}}\triangleq(\bm{a}\bm{a}^{\dagger})/\left\Vert \bm{a}\right\Vert ^{2}$,
whereas $\bm{P}_{\bm{a}}^{\perp}\triangleq(\bm{I}-\bm{P}_{\bm{a}})$
its complement.}
\end{abstract}

\begin{IEEEkeywords}
Composite Hypothesis Testing, Computational Time-Reversal, Invariant
Detection, Imaging Functions.
\end{IEEEkeywords}

\section{Problem Formulation and Related Literature}

\IEEEPARstart{T}{ime-Reversal} (TR) techniques exploit the invariance
of wave equation (in lossless and stationary media) to provide focusing
on a scattering object (or radiating source). This is achieved by
re-transmitting a time-reversed version of the scattered (or radiated)
field collected over an array and can be achieved physically \cite{Fink1993}
or synthetically \cite{Cassereau1992}. In the latter case (i.e. computational
TR, C-TR) the time-reversing procedure consists in back-propagating
numerically the field data by using the Green\textquoteright s function
of the medium in which the propagation takes place.

Accordingly, C-TR provides a powerful tool to achieve scatterer detection
and localization and represents the building rationale for many imaging
procedures in different application contexts, such as radar imaging
\cite{Odendaal1994}, subsurface prospecting \cite{Micolau2003},
through-the-wall imaging \cite{Li2010} and breast cancer detection
\cite{Hossain2013}. For this reason, theoretical limits on localization
accuracy of multiple-scatterers by means of C-TR, based on the Cramér-Rao
Lower Bound (CRLB), were obtained in \cite{Shi2007a} both for (linear)
Born Approximated (BA) and (non-linear) Foldy-Lax (FL) scattering
models.

The key entity in TR-imaging is the so-called Multistatic Data Matrix
(MDM), whose entries consist of the scattered field due to each Tx-Rx
pair for each probed frequency. Specifically, in this letter, we consider
C-TR based localization in a multi-frequency (with $L$ frequencies)
multi-static setup. We assume that $M$ point-like scatterers are
located at unknown positions $\{\bm{r}_{m}\}_{m=1}^{M}$ in $\mathbb{R}^{p}$
with unknown scattering potentials $\{\tau_{m,\ell}\}_{m=1}^{M}$
at $\ell$th angular frequency (denoted with $\omega_{\ell}$) in
$\mathbb{C}$. The Tx (resp. Rx) array consists of $N_{T}$ (resp.
$N_{R}$) isotropic point elements (resp. receivers) located at $\tilde{\bm{r}}_{i}\in\mathbb{R}^{p}$,
$i\in\{1,\ldots,N_{T}\}$ (resp. $\bar{\bm{r}}_{j}\in\mathbb{R}^{p}$,
$j\in\{1,\ldots,N_{R}\}$). The illuminators first send signals to
the probed scenario (in a known homogeneous background) and the transducer
array records the received signals. Thus, the model at $\omega_{\ell}$
is \cite{Ciuonzo2015MUSIC}:
\begin{equation}
\bm{X}_{\ell}=\bm{A}_{R,\ell}(\bm{r}_{1:M})\,\bm{M}_{\ell}(\bm{r}_{1:M},\bm{\tau}_{\ell})\,\bm{A}_{T,\ell}^{T}(\bm{r}_{1:M})+\bm{W}_{\ell},\label{eq: Time-Reversal General model}
\end{equation}
where $\bm{X}_{\ell}\in\mathbb{C}^{N_{R}\times N_{T}}$ denotes the
measured MDM at $\omega_{\ell}$ and $\bm{W}_{\ell}\in\mathbb{C}^{N_{R}\times N_{T}}$
is a noise matrix such that $\bm{w}_{\ell}\triangleq\mathrm{vec}(\bm{W}_{\ell})\sim\mathcal{C}\mathcal{N}_{N}(\bm{0}_{N},\sigma_{\ell}^{2}\,\bm{I}_{N})$,
where $N\triangleq N_{T}N_{R}$. The matrices $\bm{W}_{\ell}$, $\ell=1,\ldots L$,
are also assumed to be mutually \emph{uncorrelated}. Here $\sigma_{\ell}^{2}$
is assumed known (resp. unknown) in the non-adaptive (resp. adaptive)
case. Also, we have denoted: ($a$) the vector of scattering coefficients
at $\omega_{\ell}$ as $\bm{\tau}_{\ell}\triangleq\left[\begin{array}{ccc}
\tau_{1,\ell} & \cdots & \tau_{M,\ell}\end{array}\right]^{T}\in\mathbb{C}^{M}$; ($b$) the Tx (resp. Rx) array matrix at $\omega_{\ell}$ as $\bm{A}_{T,\ell}(\bm{r}_{1:M})\in\mathbb{C}^{N_{T}\times M}$
(resp. $\bm{A}_{R,\ell}(\bm{r}_{1:M})\in\mathbb{C}^{N_{R}\times M}$),
whose $(i,j)$th entry equals $\mathcal{G}_{\ell}(\tilde{\bm{r}}_{i},\bm{r}_{j})$
(resp. $\mathcal{G}_{\ell}(\bar{\bm{r}}_{i},\bm{r}_{j})$), where
$\mathcal{G}_{\ell}(\cdot,\cdot)$ denotes the \emph{relevant} (scalar)
background \emph{Green function at $\omega_{\ell}$} \cite{Devaney2005}.
Finally, in Eq. (\ref{eq: Time-Reversal General model}) the scattering
matrix is defined as $\bm{M}_{\ell}(\bm{r}_{1:M},\bm{\tau}_{\ell})\triangleq\mathrm{diag}(\bm{\tau}_{\ell})$
for BA model \cite{Prada1996}, whereas for FL model \cite{Shi2007a}
$\bm{M}_{\ell}(\bm{r}_{1:M},\bm{\tau}_{\ell})\triangleq\left[\mathrm{diag}^{-1}(\bm{\tau}_{\ell})-\bm{S}_{\ell}(\bm{r}_{1:M})\right]^{-1}$
holds, where the $(m,n)$th entry of $\bm{S}_{\ell}(\bm{r}_{1:M})$
equals $\mathcal{G}_{\ell}(\bm{r}_{m},\bm{r}_{n})$ when $m\neq n$
and zero otherwise.

Two popular methods for TR-imaging are the DORT \cite{Prada1996}
and the TR Multiple Signal Classification (TR-MUSIC) \cite{Devaney2005,Ciuonzo2015MUSIC,Ciuonzo2017}.
DORT method exploits the MDM spectrum so that imaging is obtained
by back-propagating each single eigenvector belonging to the \emph{signal
subspace}; this allows to selectively focus on each single scatterer
if they are well-resolved by the array. Differently, TR-MUSIC offers
a dual viewpoint with respect to (w.r.t.) DORT, and the \emph{orthogonal}
(viz. noise) \emph{subspace} is employed for imaging purposes. Both
methods, unfortunately, \emph{require knowledge of the number of scatterers
$M$ in the scene}, which is typically obtained via model-order selection
techniques \cite{Stoica2004}.

Alternatively, sub-optimal (simpler) imaging functions can be designed
based on a single scatterer model\footnote{It is worth noticing that the single-target model is coincident for
both BA and FL models, as there is no mutual interaction effects among
scatterers.}, being capable of providing an image also in the case of \emph{multiple}
\emph{scatterers} in the scene, while not requiring their exact number
$M$ (i.e. circumventing the model-order selection issue) \cite{Borcea2003,Shi2005,Shi2007}.
For this reason, in what follows, we focus at design stage on a measured
MDM (at $\omega_{\ell}$) in the form $\bm{X}_{\ell}=\bm{a}_{R,\ell}(\bm{r}_{1})\,\tau_{1,\ell}\,\bm{a}_{T,\ell}^{T}(\bm{r}_{1})+\bm{W}_{\ell}$
(i.e. the Tx and Rx array matrices collapse to column vectors) which,
after $\mathrm{vec}(\cdot)$, can be rewritten as:
\begin{equation}
\bm{x}_{\ell}=\bm{a}_{T,\ell}(\bm{r}_{1})\otimes\bm{a}_{R,\ell}(\bm{r}_{1})\,\tau_{1,\ell}+\bm{w}_{\ell}=\bm{b}_{\ell}(\bm{r}_{1})\,\tau_{1,\ell}+\bm{w}_{\ell}\,,\label{eq: single-source model (vectorization)}
\end{equation}
where $\bm{x}_{\ell}\triangleq\mathrm{vec}(\bm{X}_{\ell})\in\mathbb{C}^{N}$
and we have adopted the (short-hand) notation $\bm{b}_{\ell}(\bm{r}_{1})\triangleq\bm{a}_{T,\ell}(\bm{r}_{1})\otimes\bm{a}_{R,\ell}(\bm{r}_{1})$
(resp. $\bm{b}_{\ell}$). Based on this model, a common approach for
single ($\ell$th) frequency imaging is the so-called TR Matched Filter
(MF) \cite{Borcea2003}, formulated as:
\begin{equation}
\mathrm{I}_{\mathrm{tr}}(\bm{r},\ell)=\left|\bm{a}_{R,\ell}^{\dagger}(\bm{r})\,\bm{X}_{\ell}\,\bm{a}_{T,\ell}^{*}(\bm{r})\right|^{2}\,;\label{eq: Time-reversal MF processing}
\end{equation}
where $\bm{r}$ generically denotes the (single) scatterer \emph{probed}
location. Successively, in \cite{Shi2007} Shi and Nehorai proposed
a modification of the above imaging algorithm, based on a likelihood-maximization
inspired argument:
\begin{equation}
\mathrm{I}_{\mathrm{ml}}(\bm{r},\ell)=\left|\bm{a}_{R,\ell}^{\dagger}(\bm{r})\,\bm{X}_{\ell}\,\bm{a}_{T,\ell}^{*}(\bm{r})\right|^{2}/\left(\left\Vert \bm{a}_{R,\ell}(\bm{r})\right\Vert ^{4}\left\Vert \bm{a}_{T,\ell}(\bm{r})\right\Vert ^{4}\right)\label{eq: Nehorai Likelihood-reversal scatterer imaging}
\end{equation}
Indeed, the above imaging function can be interpreted as $\mathrm{I}_{\mathrm{ml}}(\bm{r},\ell)=\left|\hat{\tau}_{\ell}(\bm{r})\right|^{2}$,
where $\hat{\tau}_{\ell}(\bm{r})$ is the Maximum Likelihood (ML)
estimate of $\tau_{\ell}$ (assuming $\bm{r}$ \emph{known}) for the
model in (\ref{eq: single-source model (vectorization)}). Similarly,
in \cite{Shi2005} a multi-frequency (wideband) imaging algorithm,
based on true concentrated (w.r.t. $\tau_{\ell}$'s) likelihood-maximization
(termed ``likelihood imaging'') was proposed:
\begin{align}
\mathrm{I}_{\mathrm{li}}(\bm{r}) & =\prod_{\ell=1}^{L}1\,/\,\left\Vert \bm{P}_{\bm{b}_{\ell}(\bm{r})}^{\perp}\,\bm{x}_{\ell}\right\Vert ^{2}\,.\label{eq: Nehorai time-reversal likelihood imaging}
\end{align}
In this letter, we start from the same rationale as the aforementioned
works, by considering a single-source model. However, we depart from
the aforementioned approaches by constructing imaging functions based
on decision statistics which test the presence of a single scatterer
located at $\bm{r}$. Precisely, the aforementioned statistics originate
from theoretically-founded approaches for the following (composite)
hypothesis testing: 
\begin{equation}
\begin{cases}
\mathcal{H}_{0}: & \bm{x}_{\ell}=\bm{w}_{\ell},\hfill\ell=1,\ldots L\\
\mathcal{H}_{1}: & \bm{x}_{\ell}=\bm{b}_{\ell}(\bm{r})\,\tau_{\ell}+\bm{w}_{\ell},\quad\ell=1,\ldots L
\end{cases}\label{eq: hypothesis testing single-source (nominal position)}
\end{equation}
By denoting a generic statistic with $t(\bm{x}_{1:L},\bm{r})$, the
corresponding image is then formed by \emph{varying} $\bm{r}$. It
is worth remarking that the proposed approach is widely used in radar-related
applications, and it naturally arises from a cell-by-cell processing
rationale \cite{Richards2010}. One of the main contributions of this
letter is the design of imaging functions based on the well-known
Generalized Likelihood Ratio (GLR), Rao and Wald statistics \cite{Kay1998}
for the C-TR imaging problem. The scope of the present study also
includes their statistical characterization. Both adaptive ($\tau_{\ell}$
is unknown, $\sigma_{\ell}^{2}$ is a nuisance parameter) and non-adaptive
($\tau_{\ell}$ is unknown, $\sigma_{\ell}^{2}$ is known) cases will
be analyzed (as opposed to \cite{Borcea2003,Shi2005,Shi2007}), in
order to draw interesting comparisons with the imaging functions reported
in Eqs. (\ref{eq: Time-reversal MF processing}), (\ref{eq: Nehorai Likelihood-reversal scatterer imaging})
and (\ref{eq: Nehorai time-reversal likelihood imaging}), respectively.
In the adaptive case, also the Constant False-Alarm Rate (CFAR) behaviour
of the building decision statistics is thoroughly investigated, by
means of statistical invariance tools \cite{Lehmann2006}.

The rest of paper is organized as follows: Sec.~\ref{sec: Adaptive vs Non-adaptive}
tackles the imaging task as a composite hypothesis test and theoretically-founded
decision statistics are proposed; Sec.~\ref{sec: Invariance and CFARness}
analyzes their CFAR behaviour via invariance theory, while in Sec.~\ref{sec: Imaging function}
a theoretical performance analysis of the corresponding imaging functions
is provided. Furthermore, Sec. \ref{sec: Simulation results} provides
a simulation-based analysis of GLR-, Rao- and Wald-originated imaging
functions. Finally, conclusions are in Sec.~\ref{sec:Conclusions}.

\section{Non-adaptive vs. Adaptive Decision Statistics\label{sec: Adaptive vs Non-adaptive}}

In this section we develop decision statistics based on theoretically-founded
criteria which will be used as the basis for the development of corresponding
imaging functions, focusing on the specific instances of the GLR,
Rao and Wald statistics \cite{Kay1998} for the problem investigated.
With this intent, we define, for notation compactness: $\bm{\theta}_{r,\ell}\triangleq\begin{bmatrix}\Re\{\tau_{\ell}\} & \Im\{\tau_{\ell}\}\end{bmatrix}^{T}$
(unknown signal parameters at $\omega_{\ell}$); $\bm{\theta}_{s,\ell}\triangleq\sigma_{\ell}^{2}$
(nuisance parameter at $\omega_{\ell}$); $\bm{\theta}_{\ell}\triangleq\begin{bmatrix}\bm{\theta}_{r,\ell}^{T} & \bm{\theta}_{s,\ell}^{T}\end{bmatrix}^{T}$
(unknown parameters set at $\omega_{\ell}$); $\bm{\theta}_{r}\triangleq\begin{bmatrix}\bm{\theta}_{r,1}^{T} & \cdots & \bm{\theta}_{r,L}^{T}\end{bmatrix}^{T}$
(overall set of unknown signal parameters); $\bm{\theta}_{s}\triangleq\begin{bmatrix}\bm{\theta}_{s,1}^{T} & \cdots & \bm{\theta}_{s,L}^{T}\end{bmatrix}^{T}$
(overall set of nuisance parameters); $\bm{\theta}\triangleq\begin{bmatrix}\bm{\theta}_{r}^{T} & \bm{\theta}_{s}^{T}\end{bmatrix}^{T}$
(overall set of unknown parameters). Based on these definitions, in
the non-adaptive case $\bm{\theta}_{s}=\bm{\theta}_{s,\ell}=\{\emptyset\}$
and $\bm{\theta}=\bm{\theta}_{r}$ hold, respectively.

Additionally, the pdf of the $L$ MDMs is expanded as $f_{1}(\bm{x}_{1:L};\bm{\theta}_{r},\bm{\theta}_{s})=\prod_{\ell=1}^{L}f_{1}(\bm{x}_{\ell};\bm{\theta}_{\ell})$
(due to independence of noise matrices $\bm{W}_{\ell}$ among frequencies),
where:
\begin{align}
f_{1}(\bm{x}_{\ell};\bm{\theta}_{\ell}) & =(\pi\sigma_{\ell}^{2})^{-N}\,\exp\left(-\left\Vert \bm{x}_{\ell}-\bm{b}_{\ell}\,\tau_{\ell}\right\Vert ^{2}/\,\sigma_{\ell}^{2}\right),\label{eq: pdf under H1}
\end{align}
whereas the corresponding pdf under $\mathcal{H}_{0}$ is $f_{0}(\bm{x}_{1:L};\bm{\theta}_{s})=\prod_{\ell=1}^{L}f_{0}(\bm{x}_{\ell};\bm{\theta}_{s,\ell})$,
where $f_{0}(\bm{x}_{\ell};\bm{\theta}_{s,\ell})$ is obtained by
setting $\tau_{\ell}=0$ in (\ref{eq: pdf under H1}). We remark that
the composite hypothesis testing tackled in what follows is based
on assumption of a \emph{known }target\emph{ }position $\bm{r}$.
Consequently, $\bm{a}_{T,\ell}$ and $\bm{a}_{R,\ell}$ (and so $\bm{b}_{\ell}$)
are assumed to be \emph{known}. Also, given the MDMs independence
among frequencies, the GLR, Rao and Wald statistics assume the simplified
expressions \cite{Kay1998}:
\begin{gather}
\prod_{\ell=1}^{L}f_{1}(\bm{x}_{\ell};\widehat{\bm{\theta}}_{1,\ell})\,/\,f_{0}(\bm{x}_{\ell};\widehat{\bm{\theta}}_{0,s,\ell})\label{eq: GLR Rao Wald implicit}\\
\sum_{\ell=1}^{L}\left.\frac{\partial\ln f_{1}(\bm{x}_{\ell};\bm{\theta}_{\ell})}{\partial\bm{\theta}_{r,\ell}^{T}}\right|_{\bm{\theta}_{\ell}=\widehat{\bm{\theta}}_{0,\ell}}[\bm{I}_{\ell}^{-1}(\widehat{\bm{\theta}}_{0,\ell})]_{\bm{\theta}_{r,\ell},\bm{\theta}_{r,\ell}}\left.\frac{\partial\ln f_{1}(\bm{x}_{\ell};\bm{\theta}_{\ell})}{\partial\bm{\theta}_{r,\ell}}\right|_{\bm{\theta}_{\ell}=\widehat{\bm{\theta}}_{0,\ell}}\nonumber \\
\sum_{\ell=1}^{L}(\hat{\bm{\theta}}_{1,r,\ell}-\bm{\theta}_{0,r,\ell})^{T}\{[\bm{I}_{\ell}^{-1}(\widehat{\bm{\theta}}_{1,\ell})]_{\bm{\theta}_{r,\ell},\bm{\theta}_{r,\ell}}\}^{-1}\,(\hat{\bm{\theta}}_{1,r,\ell}-\bm{\theta}_{0,r,\ell})\,.\nonumber 
\end{gather}
where $\bm{I}_{\ell}(\bm{\theta})\triangleq\,\mathbb{E}\left\{ \frac{\partial\ln f_{1}(\bm{x}_{\ell};\bm{\theta}_{\ell})}{\partial\bm{\theta}_{\ell}}\frac{\partial\ln f_{1}(\bm{x}_{\ell};\bm{\theta}_{\ell})}{\partial\bm{\theta}_{\ell}^{T}}\right\} $
denotes the contribution of Fisher information matrix pertaining to
$\omega_{\ell}$ and $[\bm{I}_{\ell}^{-1}(\bm{\theta})]_{\mathrm{\bm{\theta}}_{r,\ell},\bm{\theta}_{r,\ell}}$
indicates the sub-matrix obtained by selecting from its inverse only
the elements corresponding to $\bm{\theta}_{r,\ell}$. Additionally,
we have defined $\widehat{\bm{\theta}}_{0,\ell}\triangleq\left[\begin{array}{cc}
\bm{\theta}_{0,r,\ell}^{T} & \widehat{\bm{\theta}}_{0,s,\ell}^{T}\end{array}\right]^{T}$ (resp. $\widehat{\bm{\theta}}_{1,\ell}\triangleq\left[\begin{array}{cc}
\widehat{\bm{\theta}}_{1,r,\ell}^{T} & \widehat{\bm{\theta}}_{1,s,\ell}^{T}\end{array}\right]^{T}$), with $\bm{\theta}_{0,r,\ell}\triangleq\bm{0}_{2}$ (resp. with
$\widehat{\bm{\theta}}_{1,r,\ell}$ denoting the ML estimate of $\bm{\theta}_{r,\ell}$
under $\mathcal{H}_{1}$) and $\widehat{\bm{\theta}}_{0,s,\ell}$
(resp. $\widehat{\bm{\theta}}_{1,s,\ell}$) denoting the ML estimate
of $\bm{\theta}_{s,\ell}$ under $\mathcal{H}_{0}$ (resp. under $\mathcal{H}_{1}$).

Based on this simplification and exploiting the key result provided
in the literature (e.g. \cite{Kay1998}) regarding the closed form
of the $\ell$th term of each statistic in (\ref{eq: GLR Rao Wald implicit}),
we obtain the final statistic\footnote{Indeed, it can be shown that all the three tests are statistically
equivalent to one based on $t_{\mathrm{na}}$.} $t_{\mathrm{na}}=\sum_{\ell=1}^{L}(\bm{x}_{\ell}^{\dagger}\bm{P}_{\bm{b}_{\ell}}\bm{x}_{\ell})/\sigma_{\ell}^{2}$
in the \emph{non-adaptive} case, while in the \emph{adaptive} case
it can be shown that\footnote{It is worth noticing that other well-founded decision statistics could
be considered as well for the present composite hypothesis testing.
These include the Durbin and Terrell (Gradient) statistics, as recently
employed in \cite{Ciuonzo2016}. However, it can be readily shown
(the proof is left to the reader for brevity) that in this peculiar
case they are both \emph{statistically} \emph{equivalent} to Rao test.}:
\begin{gather}
t_{\mathrm{glr}}=\prod_{\ell=1}^{L}(1+\Xi_{\ell}),\;\,t_{\mathrm{rao}}=\sum_{\ell=1}^{L}\frac{\Xi_{\ell}}{\Xi_{\ell}+1},\;\,t_{\mathrm{wald}}=\sum_{\ell=1}^{L}\Xi_{\ell},\label{eq: GLRT - Rao - Wald}
\end{gather}
where we have exploited the definitions
\begin{equation}
\Xi_{\ell}\triangleq(\bm{x}_{\ell}^{\dagger}\bm{P}_{\bm{b}_{\ell}}\bm{x}_{\ell})\,/\,(\bm{x}_{\ell}^{\dagger}\bm{P}_{\bm{b}_{\ell}}^{\perp}\bm{x}_{\ell}),\quad\ell=1,\ldots L\,.
\end{equation}
The proposed imaging functions are then obtained by varying $\bm{r}$.
Some remarks are now in order. First, it is apparent that in the non-adaptive
case the imaging function $t_{\mathrm{na}}(\bm{r})$ is a weighted
sum (via $1/\sigma_{\ell}^{2}$) of $L$ terms, each corresponding
to a single frequency. Interestingly, for $L=1$, $t_{\mathrm{na}}(\bm{r})$
assumes an expression which is comparable with the imaging functions
in (\ref{eq: Time-reversal MF processing}) and (\ref{eq: Nehorai Likelihood-reversal scatterer imaging}).
Indeed, for the generic $\omega_{\ell}$, $t_{\mathrm{na}}(\bm{r})$
simplifies to:
\begin{align}
t_{\mathrm{na}}(\bm{r}) & =\frac{(\bm{x}_{\ell}^{\dagger}\bm{P}_{\bm{b}_{\ell}(\bm{r})}\bm{x}_{\ell})}{\sigma_{\ell}^{2}}\propto\left\Vert \bm{P}_{\bm{a}_{R,\ell}(\bm{r})}\bm{X}_{\ell}\bm{P}_{\bm{a}_{T,\ell(\bm{r})}}^{T}\right\Vert _{F}^{2}\label{eq: non-adaptive imaging function}\\
 & =\left|\bm{a}_{R,\ell}^{\dagger}(\bm{r})\,\bm{X}_{\ell}\,\bm{a}_{T,\ell}^{*}(\bm{r})\right|^{2}/\left\{ \left\Vert \bm{a}_{R,\ell}(\bm{r})\right\Vert ^{2}\left\Vert \bm{a}_{T,\ell}(\bm{r})\right\Vert ^{2}\right\} ,\nonumber 
\end{align}
which enforces the use of \emph{unit-norm} Tx-Rx Green vector functions
$\bm{a}_{T,\ell}/\left\Vert \bm{a}_{T,\ell}\right\Vert $ and $\bm{a}_{R,\ell}/\left\Vert \bm{a}_{R,\ell}\right\Vert $,
as opposed to (\ref{eq: Time-reversal MF processing}) (where unnormalized
counterparts are employed) and (\ref{eq: Nehorai Likelihood-reversal scatterer imaging})
(where $\bm{a}_{T,\ell}/\left\Vert \bm{a}_{T,\ell}\right\Vert ^{2}$
and $\bm{a}_{R,\ell}/\left\Vert \bm{a}_{R,\ell}\right\Vert ^{2}$
are used instead). Such choice is in agreement with the intuitive
rationale discussed in \cite{Borcea2002}.

Secondly, it is interesting to compare the likelihood imaging in Eq.
(\ref{eq: Nehorai time-reversal likelihood imaging}) with $t_{\mathrm{glr}}$
(adaptive case), which is rewritten as:
\begin{align}
t_{\mathrm{glr}}(\bm{r}) & =\prod_{\ell=1}^{L}\left\Vert \bm{x}_{\ell}\right\Vert ^{2}/\left\Vert \bm{P}_{\bm{b}_{\ell}(\bm{r})}^{\perp}\bm{x}_{\ell}\right\Vert ^{2}\,,
\end{align}
where the sole difference observed is the introduction of the numerator
terms $\left\Vert \bm{x}_{\ell}\right\Vert ^{2}$, $\ell=1,\ldots L$.

\section{Invariance and CFARness in Adaptive Case\label{sec: Invariance and CFARness}}

When referring to the adaptive case, it is desirable to build imaging
functions which are insensitive to the unknown nuisance parameters
$\sigma_{\ell}^{2}$, $\ell=1,\ldots L$, when no scatterer is present
in the scene. To this end, we resort to the \emph{principle of invariance}
\cite{Lehmann2006} applied to the hypothesis testing problem in (\ref{eq: hypothesis testing single-source (nominal position)}),
representing an elegant way for obtaining decision statistics (resp.
tests) which ensure invariance (resp. a CFAR). To this end, hereinafter
we will search for data functions sharing invariance w.r.t. those
parameters (namely, the nuisance $\sigma_{\ell}^{2}$, $\ell=1,\ldots L$
) which are irrelevant for the decision problem considered. The preliminary
step consists in finding transformations that properly cluster data
without altering: ($i$) the structure of the hypothesis testing problem,
i.e. $\mathcal{H}_{0}\,:\,|\tau_{\ell}|=0$, $\mathcal{H}_{1}\,:\,\exists|\tau_{\ell}|>0$
($\ell=1,\ldots L$); ($ii$) the Gaussian assumption for the measured
MDMs under each hypothesis; ($iii$) the scaled-identity structure
of the noise covariance ($\sigma_{\ell}^{2}\,\bm{I}_{N}$) and the
useful signal subspace. Of course, the group invariance requirement
leads to a \emph{lossy} data reduction, with the least compression
embodied by\emph{ }the\emph{ Maximal Invariant Statistic} (MIS), organizing
the original data into equivalence classes. Hence, every invariant
statistic can be expressed in terms of the MIS \cite{Lehmann2006}.
For this reason, before proceeding, we first individuate a suitable
group fulfilling the above requirements.

To this end, we take the (vectorized) single-source model in Eq. (\ref{eq: single-source model (vectorization)})
and rewrite it in the so-called \emph{canonical form}\footnote{The canonical form representation is obtained by rotating each vector
$\bm{x}_{\ell}$ as $\bar{\bm{x}}_{\ell}\triangleq\bm{U}_{\ell}^{\dagger}\bm{x}_{\ell}$,
where $\bm{U}_{\ell}\triangleq\begin{bmatrix}\frac{\bm{b}_{\ell}}{\left\Vert \bm{b}_{\ell}\right\Vert } & \bm{U}_{c,\ell}\end{bmatrix}\in\mathcal{U}(N)$,
that is, a unitary matrix whose first column is aligned toward the
direction of $\bm{b}_{\ell}$. }:
\begin{equation}
\bar{\bm{x}}_{\ell}=\bm{e}_{1}\bar{\tau}_{\ell}+\bar{\bm{w}}_{\ell},\quad\ell=1,\ldots L,
\end{equation}
where $\bm{e}_{1}\triangleq\begin{bmatrix}1 & \bm{0}_{N-1}^{T}\end{bmatrix}^{T}$,
$\bar{\tau}_{\ell}\in\mathbb{C}$ and $\bar{\bm{w}}_{\ell}\sim\mathcal{C}\mathcal{N}_{N}(\bm{0}_{N},\sigma_{\ell}^{2}\bm{I}_{N})$.
The group of transformations leaving the hypothesis testing problem
in (\ref{eq: hypothesis testing single-source (nominal position)})
\emph{unaltered} is represented by $\mathrm{G}\triangleq\mathrm{G}_{1}\times\ldots\times\mathcal{\mathrm{G}}_{L}$,
where $\mathrm{G}_{\ell}$ is defined as follows:
\begin{align}
\mathrm{G}_{\ell} & \triangleq\{g_{\ell}\,:\,\bar{\bm{x}}_{\ell}\rightarrow\gamma_{\ell}\bm{V}_{\ell}\bar{\bm{x}}_{\ell},\quad\gamma_{\ell}\in\mathbb{R}^{+},\\
 & \,\bm{V}_{\ell}\triangleq\begin{bmatrix}e^{j\phi_{\ell}} & \bm{0}_{N-1}^{T}\\
\bm{0}_{N-1} & \bm{V}_{1,\ell}
\end{bmatrix},\,\phi_{\ell}\in(0,2\pi),\,\,\bm{V}_{1,\ell}\in\mathcal{U}(N-1)\}\nonumber 
\end{align}
After defining the partitioning $\bar{\bm{x}}_{\ell}=\begin{bmatrix}\bar{x}_{a,\ell} & \bar{\bm{x}}_{b,\ell}^{T}\end{bmatrix}^{T}$,
where $\bar{x}_{a,\ell}\in\mathbb{C}$ and $\bar{\bm{x}}_{b,\ell}\in\mathbb{C}^{N-1}$,
respectively, we are able to state the proposition providing the MIS
for the problem at hand.
\begin{prop}
The MIS for the hypothesis testing in Eq. (\ref{eq: hypothesis testing single-source (nominal position)})
under the group $\mathrm{G}$ is the $L$-dimensional vector:
\begin{gather}
\bm{t}\triangleq\begin{bmatrix}t_{1} & \cdots & t_{L}\end{bmatrix}^{T}\,,\quad t_{\ell}\triangleq\left|\bar{x}_{a,\ell}\right|^{2}/\left\Vert \bar{\bm{x}}_{b,\ell}\right\Vert ^{2}\,.\label{eq: MIS}
\end{gather}
\end{prop}
\begin{IEEEproof}
The proof is readily obtained by extending the known result developed
in the literature (see, e.g. \cite{Lehmann2006}) for the simpler
case $L=1$ and then exploiting the separability of the problem across
the frequencies $\omega_{1},\ldots\omega_{L}$.
\end{IEEEproof}
For completeness, we remark that the MIS depends on the unknown parameters
only through the corresponding induced maximal invariant \cite{Lehmann2006},
which for this specific case is $\bm{\delta}^{2}\triangleq\begin{bmatrix}\delta_{1}^{2} & \cdots & \delta_{L}^{2}\end{bmatrix}$,
where $\delta_{\ell}^{2}\triangleq\frac{\left\Vert \bm{b}_{\ell}\right\Vert ^{2}|\tau_{\ell}|^{2}}{\sigma_{\ell}^{2}}$,
corresponding to the Signal-to-Noise Ratio (SNR) experienced on $\omega_{\ell}$.

Two important considerations are now in order. First, since the MIS
in (\ref{eq: MIS}) is vector-valued \emph{no Uniformly Most Powerful
Invariant (UMPI) test exists} for the hypothesis testing in (\ref{eq: hypothesis testing single-source (nominal position)}).
Such negative result differs from that of the single-frequency ($L=1$)
case, where $t_{1}$ (i.e. a scalar-valued statistic) in Eq.~(\ref{eq: MIS})
is also the UMPI decision statistic \cite{Kay1998,Lehmann2006}.

Secondly, it can be shown (the proof is omitted for brevity) that
the equality $t_{\ell}=\Xi_{\ell}$ holds. Therefore, the MIS in Eq.~(\ref{eq: MIS})
can be rewritten as $\bm{t}=\begin{bmatrix}\Xi_{1} & \cdots & \Xi_{L}\end{bmatrix}^{T}\triangleq\bm{\Xi}$.
Then, from direct comparison of MIS with statistics in Eq. (\ref{eq: GLRT - Rao - Wald})
and, exploiting the theory in \cite{Lehmann2006}, it is readily deduced
that the tests built on the aforementioned statistics, being functions
of the original data $\bm{x}_{1:L}$ solely through the MIS $\bm{\Xi}$,
are\emph{ invariant }and therefore they all \emph{ensure} a \emph{CFAR}
w.r.t. the noise levels on the probed frequencies $\sigma_{\ell}^{2}$,
$\ell=1,\ldots L$.

Additionally, following the first consideration, since no UMPI test
exists, nothing can be said in advance on the relative performance
of the aforementioned tests. For this reason, it is useful analyzing
the structural properties of the (clairvoyant) MPI statistic\footnote{The MPI statistic is the likelihood-ratio after reduction by invariance
(i.e. based on $\bm{\Xi}$).}:
\begin{equation}
t_{\mathrm{mpi}}\triangleq\frac{f_{1}(\bm{\Xi};\bm{\delta})}{f_{0}(\bm{\Xi};\bm{\delta}=\bm{0}_{L})}=\prod_{\ell=1}^{L}\frac{f_{1}(\Xi_{\ell};\delta_{\ell})}{f_{0}(\Xi_{\ell};\delta_{\ell}=0)}\,.
\end{equation}
Clearly, since the corresponding MPI test depends on $\bm{\delta}^{2}$,
it \emph{cannot} be implemented. However, since $\Xi_{\ell}|\mathcal{H}_{1}\sim\mathcal{CF}_{1,N-1}(\delta_{\ell})$
and $\Xi_{\ell}|\mathcal{H}_{0}\sim\mathcal{CF}_{1,N}$ (these results
are given, for example, in \cite{Liu2015a}) each ratio $f_{1}(\Xi_{\ell};\delta_{\ell})/f_{0}(\Xi_{\ell};\delta_{\ell}=0)$,
$\ell=1,\ldots L$, is monotone with $\Xi_{\ell}$ (since the complex
non-central F-distribution is a totally-positive kernel of order $2$,
see \cite{Karlin1968}). Thus, it immediately follows that for each\footnote{The notation $\bm{\delta}_{1}\geq.\bm{\delta}_{2}$ means that each
element of $\bm{\delta}_{1}$ is greater or equal than the corresponding
element of $\bm{\delta}_{2}$, and at least one element of $\bm{\delta}_{1}$
is strictly greater than the corresponding element of $\bm{\delta}_{2}$.} $\bm{\delta}_{1}\geq.\bm{\delta}_{2}$ the MPI detector is an \emph{increasing
function of the vector} $\bm{\Xi}$.

Such key result allows to claim that \emph{the space of monotone functions}
of $\bm{\Xi}$ \emph{forms a complete class }of decision statistics
\cite{Brown1976}. It thus follows that the GLR, Rao and Wald statistics
in the multi-frequency case are all ``meaningful'' candidates, as
they belong to the aforementioned class (cf. Eq.~(\ref{eq: GLRT - Rao - Wald})).
Remarkably, such consideration also applies to any other statistic
built as an increasing function of $\bm{\Xi}$ (such as the geometric
or harmonic means of the elements of $\bm{\Xi}$), which also represents
a good candidate for a ``stable'' C-TR imaging.

\section{Adaptive Statistics as Imaging Procedures \label{sec: Imaging function}}

This section first provides a statistical characterization of $\bm{\Xi}$
in the presence of a mismatch of the $\bm{b}_{\ell}$'s. This preliminary
result will be useful to obtain the theoretical performance of the
proposed statistics in the \emph{realistic} case of multiple ($M>1$)
scatterers (possibly with mutual interaction effect, i.e. a FL model).
To this end, we assume a generic signal form $\bm{x}_{\ell}\sim\mathcal{N}_{\mathbb{C}}(\bm{\xi}_{\ell},\sigma_{\ell}^{2}\,\bm{I}_{N})$,
$\ell=1,\ldots L$. In this case, it holds $(\bm{x}_{\ell}^{\dagger}\bm{P}_{\bm{b}_{\ell}}\bm{x}_{\ell}/\sigma_{\ell}^{2})\sim\mathcal{C\chi}_{1}(\delta_{N,\ell})$
and $(\bm{x}_{\ell}^{\dagger}\bm{P}_{\bm{b}_{\ell}}^{\perp}\bm{x}_{\ell}/\sigma_{\ell}^{2})\sim\mathcal{C}\chi_{N-1}(\delta_{D,\ell})$,
where $\delta_{N,\ell}^{2}\triangleq(\bm{\xi}_{\ell}^{\dagger}\bm{P}_{\bm{b}_{\ell}}\bm{\xi}_{\ell})/\sigma_{\ell}^{2}$
and $\delta_{D,\ell}^{2}\triangleq(\bm{\xi}_{\ell}^{\dagger}\bm{P}_{\bm{b}_{\ell}}^{\perp}\bm{\xi}_{\ell})/\sigma_{\ell}^{2}$,
respectively. Consequently, in the mismatched case, it readily follows
that $\Xi_{\ell}\sim\mathcal{CF}_{1,N-1}(\delta_{N,\ell},\delta_{D,\ell})$. 

Based on the above result, we now characterize statistically the obtained
imaging functions for $M>1$ scatterers in the scene. Indeed, by vectorizing
the model in Eq. (\ref{eq: Time-Reversal General model}), we get
(using the short-hand notation $\bm{M}_{\ell}$ for $\bm{M}_{\ell}(\bm{r}_{1:M},\bm{\tau}_{\ell})$):
\begin{equation}
\bm{x}_{\ell}=\left[\bm{A}_{T,\ell}(\bm{r}_{1:M})\otimes\bm{A}_{R,\ell}(\bm{r}_{1:M})\right]\,\mathrm{vec}(\bm{M}_{\ell})+\bm{w}_{\ell}\,.\label{eq: Time-reversal general model vectorized}
\end{equation}
By letting $\bm{\xi}_{\ell}=\left[\bm{A}_{T,\ell}(\bm{r}_{1:M})\otimes\bm{A}_{R,\ell}(\bm{r}_{1:M})\right]\,\mathrm{vec}(\bm{M}_{\ell})$,
the mismatched analysis of Sec. \ref{sec: Invariance and CFARness}
allows to conclude that the relevant information is \emph{summarized}
through $\delta_{N,\ell}$ and $\delta_{D,\ell}$. Also, it can be
shown that for the model in Eq. (\ref{eq: Time-reversal general model vectorized}),
$\delta_{N,\ell}^{2}$ assumes the expression (when varying the \emph{probed}
location $\bm{r}$):
\begin{align}
\delta_{N,\ell}^{2}(\bm{r})=\, & \sigma_{\ell}^{-2}\,\mathrm{vec}(\bm{M}_{\ell})^{\dagger}\left[\bm{A}_{T,\ell}(\bm{r}_{1:M})\otimes\bm{A}_{R,\ell}(\bm{r}_{1:M})\right]^{\dagger}\,\bm{P}_{\bm{b}_{\ell}(\bm{r})}\nonumber \\
 & \times\left[\bm{A}_{T,\ell}(\bm{r}_{1:M})\otimes\bm{A}_{R,\ell}(\bm{r}_{1:M})\right]\,\mathrm{vec}(\bm{M}_{\ell})\,,\label{eq: delta_N,l}
\end{align}
whereas $\delta_{D,\ell}^{2}(\bm{r})$ is obtained from (\ref{eq: delta_N,l})
when replacing $\bm{P}_{\bm{b}_{\ell}(\bm{r})}$ with $\bm{P}_{\bm{b}_{\ell}(\bm{r})}^{\perp}$.
After tedious manipulations, these quantities can be rewritten in
the more intuitive form:
\begin{align}
\delta_{N,\ell}^{2}(\bm{r}) & =\,\left\Vert \bm{h}_{R,\ell}^{\dagger}(\bm{r})\,\bm{M}_{\ell}\,\bm{h}_{T,\ell}^{*}(\bm{r})\right\Vert ^{2}/\sigma_{\ell}^{2}\,;\label{eq: Noncent Num explicit}\\
\delta_{D,\ell}^{2}(\bm{r}) & =\left\{ \left\Vert \bm{A}_{R,\ell}(\bm{r}_{1:M})\,\bm{M}_{\ell}\,\bm{A}_{T,\ell}^{T}(\bm{r}_{1:M})\right\Vert _{F}^{2}/\sigma_{\ell}^{2}\right\} -\delta_{N,\ell}^{2}(\bm{r})\,;\nonumber 
\end{align}
where we have defined the vector of normalized point-spread functions
of the Tx (resp. Rx) array as $\bm{h}_{T,\ell}(\bm{r})\triangleq\{\bm{A}_{T,\ell}^{\dagger}(\bm{r}_{1:M})\bm{a}_{T,\ell}(\bm{r})\}/\left\Vert \bm{a}_{T,\ell}(\bm{r})\right\Vert ^{2}$
(resp. $\bm{h}_{R,\ell}(\bm{r})\triangleq\{\bm{A}_{R,\ell}^{\dagger}(\bm{r}_{1:M})\bm{a}_{R,\ell}(\bm{r})\}/\left\Vert \bm{a}_{R,\ell}(\bm{r})\right\Vert ^{2}$).
Consequently, it readily follows that $\Xi_{\ell}(\bm{r})\sim\mathcal{CF}_{1,N-1}(\delta_{N,\ell}(\bm{r}),\delta_{D,\ell}(\bm{r}))$,
$\ell=1,\ldots L$, and the exact pdf of the random vector $\bm{\Xi}(\bm{r})$
can be expressed as $\prod_{\ell=1}^{L}\,f_{1}(\Xi_{\ell};\delta_{N,\ell}(\bm{r}),\delta_{D,\ell}(\bm{r}))$.
Then, the pdfs of GLR, Rao and Wald imaging functions can be obtained
by \emph{transformation} of vector $\Xi(\bm{r})$, see Eq. (\ref{eq: GLRT - Rao - Wald}).
The preceding analysis is also employed to provide a theoretical characterization
of the imaging functions in Eqs.~(\ref{eq: Time-reversal MF processing}-\ref{eq: Nehorai Likelihood-reversal scatterer imaging}-\ref{eq: Nehorai time-reversal likelihood imaging}-\ref{eq: non-adaptive imaging function}),
as shown hereinafter.

\emph{MF imaging}: The function $\mathrm{I}_{\mathrm{tr}}(\bm{r},\ell)$
in (\ref{eq: Time-reversal MF processing}) can be rewritten by exploiting
$\bm{a}_{R,\ell}^{\dagger}(\bm{r})\,\bm{X}_{\ell}\,\bm{a}_{T,\ell}^{*}(\bm{r})=\bm{b}_{\ell}^{\dagger}(\bm{r})\,\bm{x}_{\ell}$
(achieved with the use of $\mathrm{vec}(\cdot)$ and Kronecker product
properties), thus leading to $\mathrm{I}_{\mathrm{tr}}(\bm{r},\ell)/(\left\Vert \bm{b}_{\ell}(\bm{r})\right\Vert ^{2}\sigma_{\ell}^{2})\sim\mathcal{C\chi}_{1}(\delta_{N,\ell}(\bm{r}))$,
which in turn provides
\begin{equation}
\mathrm{I}_{\mathrm{tr}}(\bm{r},\ell)\sim\mathcal{C\chi}_{1}\left(\delta_{N,\ell}(\bm{r})\,,\,\sigma_{\ell}^{2}\left\Vert \bm{b}_{\ell}(\bm{r})\right\Vert ^{2}\right)\,.
\end{equation}
\emph{ML imaging}: By similar reasoning as MF imaging, $\mathrm{I}_{\mathrm{ml}}(\bm{r},\ell)$
in (\ref{eq: Nehorai Likelihood-reversal scatterer imaging}) is distributed
as (recall that $\left\Vert \bm{b}_{\ell}(\bm{r})\right\Vert ^{2}=\left\Vert \bm{a}_{T,\ell}(\bm{r})\right\Vert ^{2}\left\Vert \bm{a}_{R,\ell}(\bm{r})\right\Vert ^{2}$):
\begin{equation}
\mathrm{I}_{\mathrm{ml}}(\bm{r},\ell)=\mathcal{C\chi}_{1}\left(\delta_{N,\ell}(\bm{r})\,,\,\sigma_{\ell}^{2}\,/\,\left\Vert \bm{b}_{\ell}(\bm{r})\right\Vert ^{2}\right)\,.
\end{equation}
\emph{Likelihood imaging:} The imaging function $\mathrm{I}_{\mathrm{li}}(\bm{r})$
in Eq.~(\ref{eq: Nehorai time-reversal likelihood imaging}) can
be rewritten as $\mathrm{I}_{\mathrm{li}}(\bm{r})=(\prod_{\ell=1}^{L}\sigma_{\ell}^{2}\,f_{\ell})^{-1}$,
where $f_{\ell}\sim\mathcal{C\chi}_{N-1}(\delta_{D,\ell}(\bm{r}))$,
$\ell=1,\ldots L$. 

\emph{Non-adaptive imaging:} It can be easily shown that $t_{\mathrm{na}}(\bm{r})$
in (\ref{eq: non-adaptive imaging function}) is distributed as $t_{\mathrm{na}}(\bm{r})\sim\mathcal{C\chi}_{1}(\delta_{N,\ell}(\bm{r}))$.
We recall that such function requires the knowledge of $\sigma_{\ell}^{2}$,
$\ell=1,\ldots L$, which may be not known in an adaptive scenario.

\section{Simulation Analysis \label{sec: Simulation results}}

In this section we focus on 2-D localization in a homogeneous background,
where the relevant (scalar) background Green\emph{ }function\emph{
at $\omega_{\ell}$} is\footnote{We discard the irrelevant constant term $j/4$.}
$\mathcal{G}_{\ell}(\bm{x}',\bm{x})=H_{0}^{(1)}\left(\kappa_{\ell}\left\Vert \bm{x}'-\bm{x}\right\Vert \right)$,
\cite{Devaney2005}. Here $H_{n}^{(1)}(\cdot)$ and $\kappa_{\ell}=2\pi/\lambda_{\ell}$
denote the $n$th order \emph{Hankel} function of the 1st kind and
the wavenumber ($\lambda_{\ell}$ is the wavelength corresponding
to $\omega_{\ell}$), respectively.

\begin{figure}[t]
\centering{}\includegraphics[width=0.9\columnwidth]{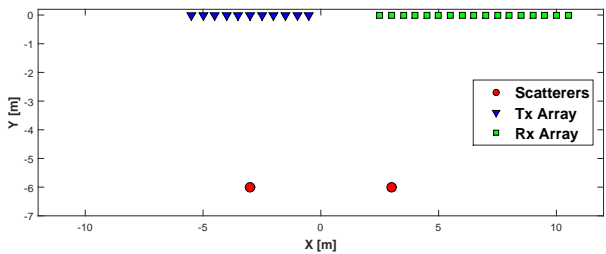}\caption{Geometry for the considered imaging problem in 2-D space.\label{fig: Setup non-colocated}}
\end{figure}

In what follows, we consider a non-colocated setup with half-meter
spaced Tx/Rx arrays displaced as shown in Fig.~\ref{fig: Setup non-colocated}
($N_{T}=11$ and $N_{R}=17$, in blue ``$\bigtriangledown$'' and
green ``$\square$'' markers, respectively). For this scenario,
we consider $L=3$ frequencies for active probing, with considered
frequencies equal $\left(f_{1},f_{2},f_{3}\right)=\left(300,600,900\right)\,\mathrm{MHz}$
(recall that $f_{\ell}=\omega_{\ell}/2\pi$), corresponding to $\left(\lambda_{1},\lambda_{2},\lambda_{3}\right)=\left(1,0.5,0.333\right)\,\mathrm{m}$.
For this example, the noise variance levels $\sigma_{\ell}^{2}$ pertaining
to the considered frequencies are set to $\left(\sigma_{1}^{2},\sigma_{2}^{2},\sigma_{3}^{2}\right)=\left(-15,-5-15\,\right)\mathrm{dB}$. 

For simplicity, we consider $M=2$ targets in the considered area,
located at $\bm{x}_{1}=\left[\begin{array}{cc}
-1 & -6\end{array}\right]^{T}\,\mathrm{m}$ and $\bm{x}_{2}=\left[\begin{array}{cc}
+1 & -6\end{array}\right]^{T}\,\mathrm{m}$ (reported as red ``$\circ$'' markers in Fig.~\ref{fig: Setup non-colocated})
and having scattering coefficients $\bm{\tau}_{\ell}=\left[\begin{array}{cc}
3 & 4\end{array}\right]^{T}$. For the aforementioned scenario, we report the averaged (over $100$
independent runs) spectrum of the imaging functions ($i$) $\log\left[t_{\mathrm{glr}}(\bm{r})\right]$,
($ii$) $t_{\mathrm{rao}}(\bm{r})$, ($iii$) $t_{\mathrm{wald}}(\bm{r})$
and ($iv$) $\log\left[\mathrm{I}_{\mathrm{li}}(\bm{r})\right]$ being
proposed/investigated in this letter. We remark that, aiming at a
fair comparison, GLR- and likelihood-imaging have been reported in
their log-versions, so as to compare all the imaging functions in
terms of sums of contributions over the considered frequencies. 

\begin{figure*}
\centering{}\includegraphics[width=0.9\paperwidth]{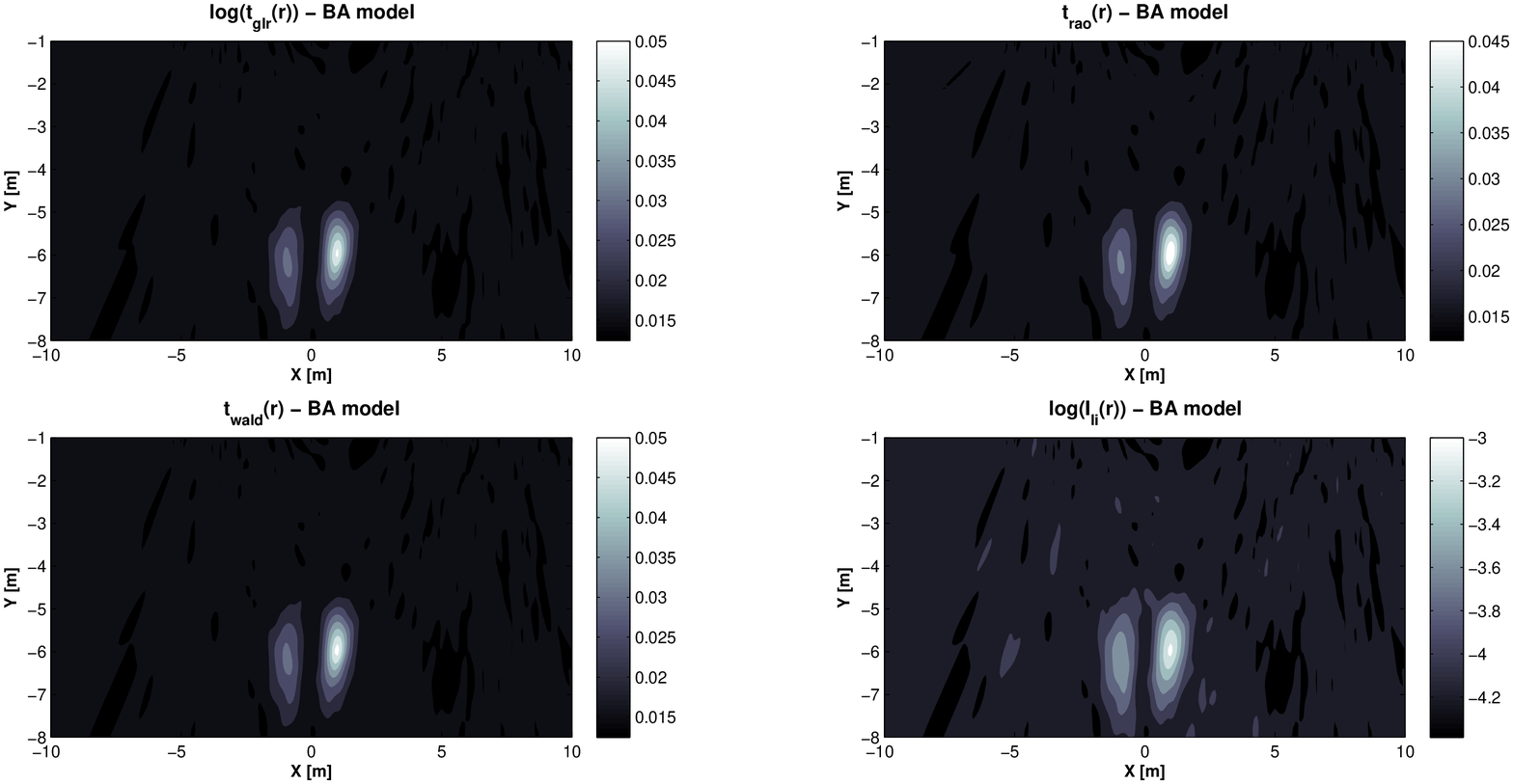}\caption{Comparison of considered imaging functions; BA scattering model. Scatterers
are located at $\bm{x}_{1}=\left[\protect\begin{array}{cc}
-1 & -6\protect\end{array}\right]^{T}\,\mathrm{m}$ and $\bm{x}_{2}=\left[\protect\begin{array}{cc}
+1 & -6\protect\end{array}\right]^{T}\,\mathrm{m}$. \label{fig: Imaging Functions BA model}}
\end{figure*}

\begin{figure*}
\centering{}\includegraphics[width=0.9\paperwidth]{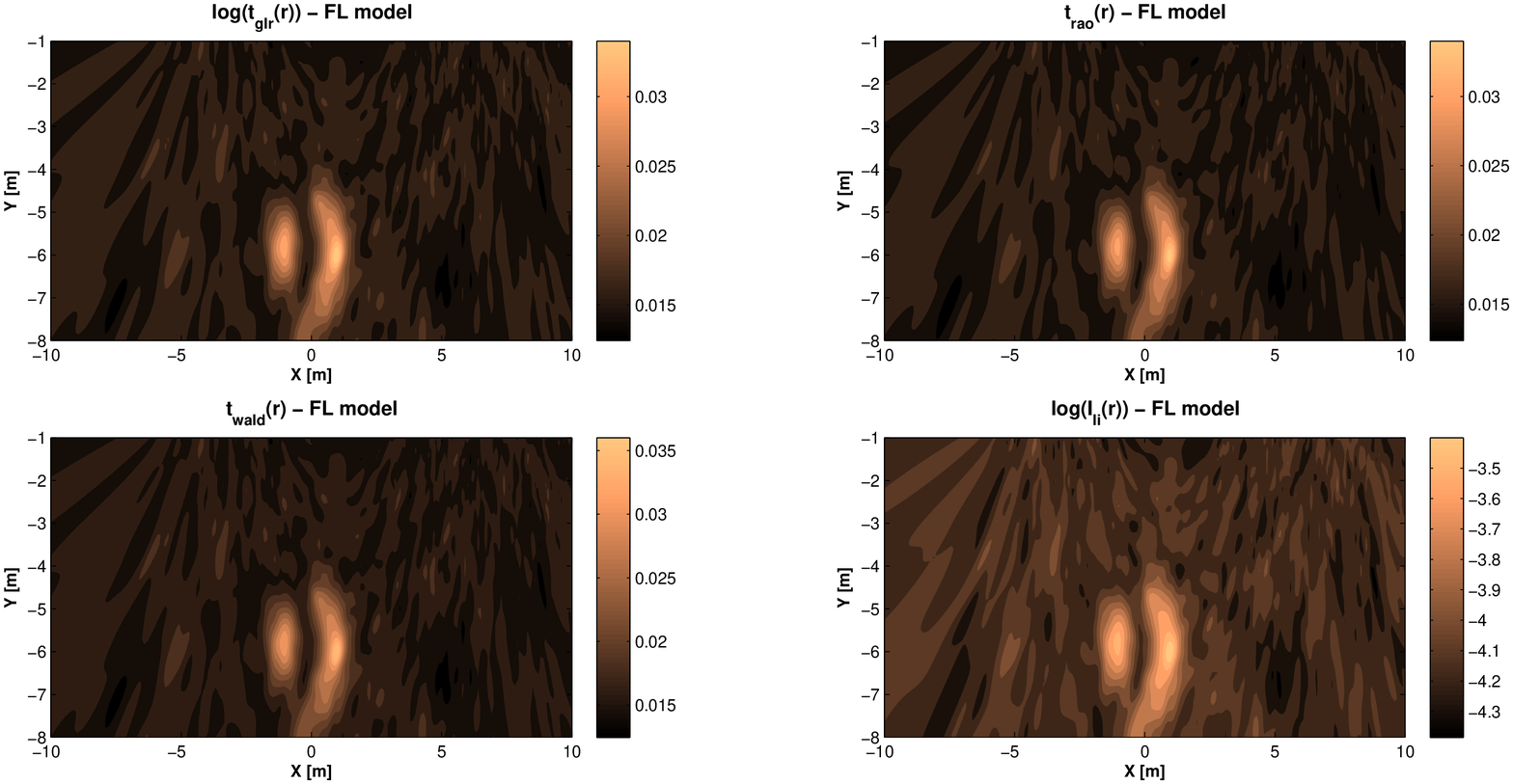}\caption{Comparison of considered imaging functions; FL scattering model. Scatterers
are located at $\bm{x}_{1}=\left[\protect\begin{array}{cc}
-1 & -6\protect\end{array}\right]^{T}\,\mathrm{m}$ and $\bm{x}_{2}=\left[\protect\begin{array}{cc}
+1 & -6\protect\end{array}\right]^{T}\,\mathrm{m}$.\label{fig: Imaging Functions FL model}}
\end{figure*}

In Fig.~\ref{fig: Imaging Functions BA model} we show the results
corresponding to BA scattering model, whereas the corresponding results
for FL model are reported in Fig.~\ref{fig: Imaging Functions FL model}.
From inspection of both the figures, it is apparent that the proposed
imaging functions (being based on statistical testing and designed
to enjoy weak-sensitivity to $\sigma_{\ell}^{2}$s) offer an improved
contrast in resolving the two scatterers considered in comparison
to likelihood-imaging. Those imaging functions are also observed to
exhibit a more stable behaviour with a varying level of noise power.
Also, they are observed to perform all equally well, as apparent from
Figs.~\ref{fig: Imaging Functions BA model} and \ref{fig: Imaging Functions FL model},
respectively. Indeed, their relative performance varies from case
to case and reflects non-optimality of all the considered testing
procedures for finite samples, see \cite{Kay1998}.

Finally, useful considerations can be drawn on the relative performance
in both the cases of BA (Fig.~\ref{fig: Imaging Functions BA model})
and FL (Fig.~\ref{fig: Imaging Functions FL model}) scattering models.
Indeed, in the case of FL scattering, a higher distortion is generally
observed w.r.t. BA case. This is mainly due to the non-linearity of
the scattering model, which is not accounted by all the considered
procedures, being based on a single-target model assumption (and thus
not reflecting mutual interaction effects).


\section{Conclusions\label{sec:Conclusions}}

In this letter imaging functions for wideband C-TR have been devised
based on GLR, Rao and Wald statistics under the single-source model.
Both non-adaptive and adaptive (where a supporting CFAR analysis through
invariance principle has been provided) have been analyzed. The proposed
imaging functions have been also compared with other imaging functions
proposed in the literature. For all these functions a theoretical
characterization for the multiple-scatterers case (possibly with mutual
interaction) has been derived and shown to depend only on the non-centrality
parameter functions $\delta_{N,\ell}(\bm{r})$ and $\delta_{D,\ell}(\bm{r})$
(Eq.~(\ref{eq: Noncent Num explicit})).

\bibliographystyle{IEEEtran}
\bibliography{IEEEabrv,imaging_refs,bib_adapt_det}

\end{document}